# Fake News Identification on Twitter with Hybrid CNN and RNN Models

**Oluwaseun Ajao**
C3Ri Research Institute
Sheffield Hallam University
United Kingdom
oajao@acm.org

**Deepayan Bhowmik**
C3Ri Research Institute
Sheffield Hallam University
United Kingdom
d.bhowmik@ieee.org

**Shahrzad Zargari**
C3Ri Research Institute
Sheffield Hallam University
United Kingdom
s.zargari@shu.ac.uk

## ABSTRACT

The problem associated with the propagation of fake news continues to grow at an alarming scale. This trend has generated much interest from politics to academia and industry alike. We propose a framework that detects and classifies fake news messages from Twitter posts using hybrid of convolutional neural networks and long-short term recurrent neural network models. The proposed work using this deep learning approach achieves 82% accuracy. Our approach intuitively identifies relevant features associated with fake news stories without previous knowledge of the domain.

## 1 INTRODUCTION

The growing influence experienced by the propaganda of fake news is now cause for concern for all walks of life. Election results are argued on some occasions to have been manipulated through the circulation of unfounded and doctored stories on social media including microblogs such as Twitter. All over the world, the growing influence of fake news is felt on daily basis from politics to education and financial markets. This has continually become a cause of concern for politicians and citizens alike. For example, on April 23rd 2013, the Twitter account of the news agency, Associated Press, which had almost 2 million followers at the time was hacked.

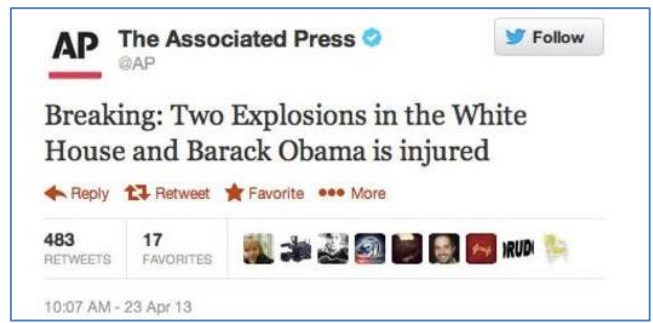

**Figure 1: Tweet allegedly sent by the Syrian Electronic Army from hacked Twitter account of Associated Press**

The impact may also be severe. The following message was sent, "Breaking: Two Explosions in the White House and Barack Obama is injured" (shown in Figure 1). This message led to a flash crash on the New York Stock Exchange, where investors lost 136 billion dollars on the Standard & Poors Index in two minutes [1]. It would be interesting and indeed beneficial if the origin of messages could be verified and filtered where the fake messages were separated from authentic ones. The information that people listen to and share in social media is largely influenced by the social circles and relationships they form online [2]. The feat of accurately tracking the spread of fake messages and especially news content would be of interest to researchers, politicians, citizens as well as individuals all around the world. This can be achieved by using effective and relevant "social sensor tools" [3]. This need is more important in countries that have trusted and embraced technology as part of their electoral process and thus adopted e-voting. [4] For example, in France and Italy, even though internet users may not accurately represent the demographics of the entire population, opinions on social media and mass surveys of citizens are correlated. They are both found to be largely influenced by external factors such as news stories from newspapers, TV and ultimately on social media.

Table 1: Most Circulated and Engaging Fake News Stories on Facebook in 2016

| S/N | Fake News Headline | Category |
|---|---|---|
| 1 | Obama Signs Executive Order Banning The Pledge of Allegiance in Schools Nationwide | Politics |
| 2 | Woman arrested for defecating on boss' desk after winning the lottery | Crime |
| 3 | Pope Francis Shocks World, Endorses Donald Trump for President, Releases Statement | Politics |
| 4 | Trump Offering Free One-Way Tickets to Africa & Mexico for Those Who Wanna Leave America | Politics |
| 5 | Cinnamon Roll Can Explodes Inside Man's Butt During Shoplifting Incident | Crime |
| 6 | Florida Man dies in meth-lab explosion after lighting farts on fire | Crime |
| 7 | FBI Agent Suspected in Hillary email Leaks Found Dead in Apparent Murder-Suicide | Politics |
| 8 | RAGE AGAINST THE MACHINE To Reunite And Release Anti Donald Trump Album | Politics |
| 9 | Police Find 19 White Female Bodies In Freezers With "Black Lives Matter" Carved Into Skin | Crime |
| 10 | ISIS Leader calls for American Muslim Voters to Support Hillary Clinton | Politics |

In addition, there is a growing and alarming use of social media for anti-social behaviors such as cyberbullying, propaganda, hate crimes, and for radicalization and recruitment of individuals into terrorism organizations such as ISIS [5]. A study by Burgess et al. into the top 50 most retweeted stories with pictures of the Hurricane Sandy disaster found that less than 25% were real while the rest were either fake or from unsubstantiated sources [6]. Facebook announced the use of "filters" for removing hoaxes and fake news from the news feed on the world's largest social media platform [7]. The development followed concerns that the spread of fake news on the platform might have helped Donald Trump win the US presidential elections held in 2016 [8]. According to the social media site, 46% of the top fake news stories circulated on Facebook were about US politics and election [9]. Table 1 gives detail of the top ranking news stories that were circulated on Facebook in year 2016.

Tambuscio et al proposed a model for tracking the spread of hoaxes using four parameters: spreading rate, gullibility, probability to verify a hoax, and forgetting one's current belief[10]. Many organizations now employ social media accounts on Twitter, Facebook and Instagram for the purposes of announcing corporate information such as earnings reports and new product releases. Consumers, investors, and other stake holders take these news messages as seriously as they would for any other mass media [11]. Other reasons that fake news have been widely proliferated include for humor, and as ploys to get readers to click on sponsored content (also referred to as "clickbait").

In this work, we aim to answer the following research questions:
- Given tweets about a news item or story, can we determine their truth or authenticity based on the content of the messages?
- Can semantic features or linguistic characteristics associated with a fake news story on Twitter be automatically identified without prior knowledge of the domain or news topic?

The remainder of the paper is structured as follows: in Section 2, we discuss related works in rumor and fake news detection, Section 3 presents the methodology that we propose and utilize in our task of fake news detection, while the results of our findings are discussed and evaluated in Section 4. Finally, conclusions and future work are presented in Section 5.

## 2  RELATED WORKS

The work on fake news detection has been initially reviewed by several authors, who in the past referred to it only "rumors", until 2016 and the US presidential election. During the election, the phrase "fake news" became popular with then-candidate, Donald Trump. Until recently, Twitter allowed their users to communicate with 140 characters on its platform, limiting users ability to communicate with others. Therefore, users who propagate fake news, rumors and questionable posts have been found to incorporate other mediums to make their messages go viral. A good example happened in the aftermath of Hurricane Sandy, where enormous amounts of fake and altered images were circulating on the internet. Gupta et al used a Decision Tree classifier to distinguish between fake and real images posted about the event[12].

Neural networks are a form of machine learning method that have been found to exhibit high accuracy and precision in clustering and classification of text [13]. They also prove effective in the prompt detection of spatio-temporal trends in content propagation on social media. In this approach, we combine this with the efficiency of recurrent neural networks (RNN) in the detection and semantic interpretation of images. Although this hybrid approach in semantic interpretation of text and images is not new [14][15], to the best of our knowledge, this is the first attempt involving the use of a hybrid approach in the detection of the origin and propagation of fake news posts. Kwon et al identified and utilized three hand-crafted feature types associated with rumor classification including: (1) Temporal features - how a tweet propagates from one time window to another; (2) Structural Features - how the influence or followership of posters affect other posts; and (3) Linguistic Features - sentiment categories of words[16].

Previous work done by Ferrara achieved 97% accuracy in detecting fake images from tweets posted during the Hurricane Sandy disaster in the United States They performed a characterization analysis, to understand the temporal, social reputation and influence patterns for the spread of fake images by examining more than 10,000 images posted on Twitter during the incident[12]. They used two broad types of features in the detection of fake images posted during the event. These include seven user-based features such as age of the user account, followers' size and the follower-followee ratio. They also deduced eighteen tweet-based features such as tweet length, retweet count, presence of emoticons and exclamation marks.

Aggarwal et al had identified four certain features based on URLs, protocols to query databases content and followers networks of tweets associated with phishing, which present a similar problem to fake and non-credible tweets but in their case also have the potential to cause significant financial harm to someone clicking on the links associated with these "phishing" messages[17].

Yardi et al developed three feature types for spam detection on Twitter; which includes searches for URLs, matching username patterns, and detection of keywords from supposedly spam messages[18]. O'Donovan et al identified the most useful indicators of credible and non-credible tweets as URLs, mentions, retweets, and tweet lengths[19].

Other works on the credibility and veracity identification on Twitter include Gupta et al that developed a framework and real-time assessment system for validating authors content on Twitter as they are being posted[20]. Their approach assigns a graduated credibility score or rank to each tweet as they are posted live on the social network platform.

## 3 METHODOLOGY

The approach of our work is twofold. First is the automatic identification of features within Twitter post without prior knowledge of the subject domain or topic of discussion using the application of a hybrid deep learning model of LSTM and CNN models. Second is the determination and classification of fake news posts on Twitter using both text and images.

We posit that since the use of deep learning models enables automatic feature extraction, the dependencies amongst the words in fake messages can be identified automatically [13] without expressly defining them in the network. The knowledge of the news topic or domain being discussed would not be necessary to achieve the feat of fake news detection.

### 3.1 The Deep learning Architectures

We implemented three deep neural network variants. The models applied to train the datasets include:

3.1.1 Long-Short Term Memory (LSTM)

LSTM recurrent neural network (RNN) was adopted for the sequence classification of the data. The LSTM [7] remains a popular method for the deep learning classification involving text since when they first appeared 20 years ago [22]

3.1.2  LSTM with dropout regularization

LSTM with dropout regularization [23] layers between the word embedding layer and the LSTM layer was adopted to avoid over-fitting to the training dataset. Following this approach, we randomly selected and dropped weights amounting to 20% of neurons in the LSTM layer.

*3.1.3 LSTM with convolutional neural networks (CNN)*
We included a 1D CNN [14] immediately after the word embedding layer of the LSTM model. We further added a max pooling layer to reduce dimensionality of the input layer while preserving the depth and avoid over-fitting of the training data. This also helps in reducing computational time and resources in the training of the model. The overall aim is to ultimately improve model prediction accuracy.

### 3.2 About the Dataset

The dataset consisted of approximately 5,800 tweets centered on five rumor stories. The tweets were collected and used in the works by Zubiaga et al[24. The dataset consisted of original tweets and they were labeled as rumor and non-rumors. The events were widely reported in online, print and conventional electronic media such radio and Television at the time of occurrence:

- CharlieHebdo
- SydneySiege
- Ottawa Shooting
- Germanwings-Crash
- Ferguson Shooting

We applied ten-fold cross validation on the entire dataset of 5,800 tweets and performed padding of the tweets (i.e., adding zeros to the tweets for uniform inclusion in the feature vector for analysis and processing).

### 3.3 Recurrent Neural Network (RNNs)

This type of neural network has been shown to be effective in time and sequence based predictions [25]. Twitter posts can be likened to events that occur in time [16] where the intervals between the retweet of one user to another is contained within a time window and treated in sequential modes. Rumors have been examined in the context of varying time windows [26]. Recurrent Neural Networks were initially limited by the problem associated with the adjustment of weights over time. Several methods have been adopted in solving the vanishing gradient problem but can largely be categorized into two types, namely, the exploding gradient and the vanishing gradient. Solutions adopted for the former include truncated back propagation, penalties and gradient clipping (these resolve the exploding gradient problem), while the vanishing gradient problem has been resolved using dynamic weight initializations, the echo state networks (ESN) and Long-Short Term Memory (LSTMs). LSTMs will be the main focus of this work as they preserve the memory from the last phase and incorporate this in the prediction task of the neural network model. Weights are the long term memories of the neural network.

### 3.4 Incorporating Convolutional Neural Network

Another popular model is the convolutional neural network (CNN) which has been well known for its application in image processing as well as use in text mining [27]. We posit that addition of the hybrid method would improve performance of the model and give much better results for the content based fake news detection. However, the hybrid implementation for this work so far involves a text-only approach.

### 3.5 Selection of Training Parameters

The following hyper-parameters were optimized using a grid search approach and optimal values derived for the following batch size, epochs, learning rates, activation function and dropout regularization rate which is set at 20%. This hybrid method of detecting fake news using this approach complements the natural language semantic processing of text by ensuring that images allow for the disambiguation of posts and making them more enriched in the identification repository. We posit that it would also be interesting to look at users' impact in the propagation of these messages and fake news content on Twitter.

The aim of the study is to detect the veracity of posts on Twitter. Possible applications include assisting law enforcement agencies in curtailing the spread and propagation of such messages, especially when they have negative implications and consequences for readers who believe them.

## 4 EVALUATION, RESULTS AND DISCUSSION

Our deep learning model intuitively achieves 82% accuracy on the classification task in detecting fake news posts without prior domain knowledge of the topics being discussed. So far in the experiments completed, it is revealed that the plain vanilla LSTM model achieved the best performance in terms of precision, recall and FMeasure and an accuracy of 82% as shown in Table 2. On the other hand, the LSTM method with dropout regularization performed the least in terms of the metrics adopted. This is likely a result of under fitting the model, in addition to the lack of sufficient training data and examples within the network. Another reason for the low performance of the dropout regularization may be the depth of the network, as it is relatively shallow, resulting in the drop-out layer being quite close to the input and output layers of the model. This could severely degrade the performance of the method. An alternative to improve model performance could be through Batch Normalization [28], where the input values in the layers have a mean activation of zero and standard deviation of one as in a standard normal distribution; this is beyond the scope of this current work.

The LSTM-CNN hybrid model performed better than the dropout regularization model, with a 74% accuracy and an FMeasure of 39.7%. However insufficient training examples for the neural network model led to negative appreciation against the plain-vanilla LSTM model.

The precision of 68% achieved by the state of the art PHEME dataset by [24] was still higher than the results obtained so far. However we believe that with the inclusion of more training data from the reactions to the original twitter posts, there will be more significant improvement in model performance.

## 5 CONCLUSION AND FUTURE WORK

We have presented a novel approach for the detection of fake news spread on Twitter from message text and images. We achieved an 82% accuracy performance beating the state of the art on the PHEME Dataset. It is expected that our approach will achieve much better results following incorporation of fake image disambiguation (found in these tweets aimed at making the posts go viral).

We leverage on a hybrid implementation of two deep learning models to find that we do not require a very large number of tweets about an event to determine the veracity or credibility of the messages. Our approach gives a boost in the achievement of a higher performance while not requiring a large amount of training data typically associated with deep learning models. We are progressively examining the inference of the tweet geo-locations and origin of these fake news items and the authors that propagate them. It would be interesting if also the training data required in this task was relatively smaller such that fake news items can be quickly identified. It would be further beneficial if the origin and location [29] of fake posts were tracked with minimal computational resources.

Table 2: Proposed Deep Learning Methods in Fake News Detection

| Technique | ACC | PRE | REC | F-M |
| --- | --- | --- | --- | --- |
| LSTM | 82.29 | 44.35 | 40.55 | 40.59 |
| LSTMDrop | 73.78 | 39.67 | 29.71 | 30.93 |
| LSTM-CNN | 80.38 | 43.94 | 39.53 | 39.70 |

Deep learning models such as CNN and RNN often require much larger datasets, and in some cases multiple layers of neural networks for the effective training of their models. In our case, we have a small dataset of 5,800 tweets. In our ongoing and future work we have collected the reactions of other users to these messages via the Twitter API in the magnitude of hundreds of thousands, with the aim of enriching the size of the training dataset and thus improving the robustness of the model performance. We expect that this will also help to draw more actionable insights for the propagation of these messages from one user to another and how they react—specifically if they embraced or refrained from becoming evangelists and promoters of these messages to other users on the platform.